\documentclass[12pt]{iopart}
\usepackage{epsf,epsfig,latexsym}

\begin{document}

%\title[Author guidelines for IOPP journals]{Kaon Phase Space Density}
\title{Kaon Phase Space Density in Heavy Ion Collisions}
\author{Michael Murray}
\address{Cyclotron Institute, Texas A\&M University, College Station,
TX 77843-3366, m-murray@tamu.edu}

\begin{abstract}
The first measurement of kaon phase space densities are presented as
a function of $m_T$,
 $\sqrt{s_{nn}}$ and the number of participants.
The kaon phase space density increases with the number
of participants from $e^+e^-$ to $Pb+Pb$ collisions. % but falls with  $\sqrt{s_{nn}}$ .
However the ratio of the kaon and pion phase space
densities at low $P_T$ is independent of
number of participants for $\sqrt{s_{nn}}=17$GeV.

    This paper is dedicated to
Francis Riccardelli, engineer for the Port Authority, who died on September 11th 2001 while evacuating
 others.
\end{abstract}



%\maketitle

\section{Introduction}
A particle's phase space density is defined as
\begin{equation}
 f({\bf p}, {\bf x}) \equiv \frac{(2\pi \hbar c)^3}{(2s+1)}
  \frac{d^6N}{dp^3dx^3}
 \label{eq:fdefine}
\end{equation}
where s is the particle's spin. A way to measure the average value of $f(\bf p)$
over the volume of the system sampled by interferometry measurements was first
suggested by Bertsch \cite{BERTSCH}.
For a system in chemical  equilibrium at a temperature  $T$ and chemical potential $\mu$
\begin{equation}
 f(E) = \frac{1}{e^{(E-\mu)/T}\pm 1}
 \label{eq:fbose}
\end{equation}
where E is the energy and $\pm 1$ selects bosons or fermions.
Since the energy E may depend on the transverse flow in the system the
phase space density is sensitive to both the dynamics and thermodynamics of the
system. The pion phase space density was first measured
, as a function of $m_T$, by Ference  {\it et al.},
 \cite{FERENC}. For Pb+Pb collisions at   center of mass energy $\sqrt{s_{nn}}=17$GeV a temperature
 of 120MeV was extracted as well as a measure of the average transverse flow.
Later measurements of proton, antiproton and pion phase densities showed that
they also depend on the number of participants and
 $\sqrt{s_{nn}}$, \cite{Murray}.
In this paper we derive phase space densities for $k^\pm$ versus $m_T$,
 $\sqrt{s_{nn}}$ and the number of participants and compare to   $\pi^\pm$, $p$
 and $\bar p$  densities from   \cite{Murray}.

%\section{Formalism}
For a dilute system, {\it i.e.} $f \ll 1$,
Eqn.~\ref{eq:fbose} gives
\begin{equation}
 f_k  \approx  e^{\mu -E_k/T} %= e^{\mu_u-\mu_s -E_k/T}
\end{equation}
Since $E=m_T cosh(y)$
we would expect $f_k$ to be exponential in $m_T$ with a slope that depends
on the strength of the transverse flow.
If the relevant degrees of freedom are quarks then since
 the $k^+$ contains both $u$ and ${\bar s}$ quarks
\begin{eqnarray}
\label{eq:fquark}
f_{k^+}  &\approx  & e^{\mu_u-\mu_s -E_k/T}  \\
f_{k^-}  & \approx  & e^{-\mu_u+\mu_s -E_k/T}
\end{eqnarray}
This implies that
\begin{equation}
\label{eq:fkmdfkp}
\frac{f_{k^-}}{f_{k^+}} = e^{-2\mu_u+2\mu_s}
\end{equation}
 and so $\frac{f_{k^-}}{f_{k^+}}$ can give information on the light and strange quark
 chemical potentials.

 For pions we would expect the chemical potential of $u$ and ${\bar u}$  quarks
 to approximately cancel  and so
\begin{equation}
\label{eq:fpi}
f_{\pi}  \approx   e^{ -E_k/T}
\end{equation}
and
\begin{equation}
\label{eq:fkpi}
f_{k^+}/f_{\pi} \approx e^{\mu_u-\mu_s}
\end{equation}

Thus the ratio of phase densities $ f_{k^+}/f_{\pi}$ is controlled by the balance of the
light and strange quark chemical potentials.

\section{Method}
NA44 has measured the kaon source size in
3 dimensions with HBT,
as well as single particle
spectra \cite{NA44KHbt,NA44ksng}.
Dividing %$\sqrt{\lambda}
$d^3N_k/dp^3$ by the Lorentz invariant volume,
\cite{BERTSCH,WEID99A}
%\cite{BERTSCH,E87797,WEID99A}.
gives  the spatially averaged
   phase space density.
\begin{equation}
 \langle f_k \rangle  =
\frac{\pi^{\frac{3}{2}} (\hbar c)^3}{(2s+1)}
{d^3N_k\over dp^3} \frac{1}{R_{side}\sqrt{R^2_{out}
R^2_{long} - R^4_{outlong}}} .
 \label{eq:pifaze}
\end{equation}
 At y=0   $R_{outlong}$ should be identically zero
  \cite{CHAPMAN} and so we ignore it
 in this paper since the data are close to mid-rapidity.

 \section{Results}
\begin{figure}
\epsfig{file=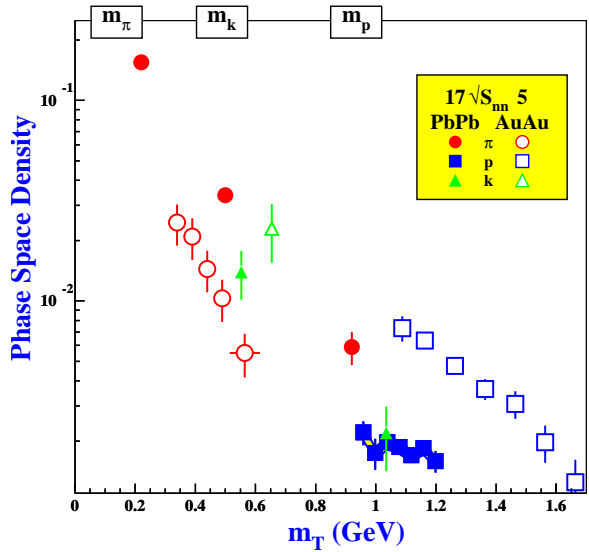,width=8.5cm,angle=0}
\caption{\label{fg:rfaze}  $\pi^+$, proton and $k^+$ phase space densities versus
$m_T$ at  $\sqrt{s_{nn}}=17$GeV, solid circles, square and diamonds.
The pion point at $m_T=0.9$ (GeV) is actually  $\pi^-$. The open
symbols are for $\sqrt{s_{nn}}=5$GeV.}
\end{figure}

 Figure~\ref{fg:rfaze} shows kaon, pion and proton phase space densities
versus the transverse mass $m_T$ for Au+Au and Pb+Pb
collisions at $\sqrt{s_{nn}}=$ 5 and 17GeV respectively.
 Although all phase densities drop with      $m_T$
the slope is flatter for protons than for pions and kaons as might be expected
from the effect of transverse flow.
The kaon and pion slopes for  $\sqrt{s_{nn}}=17$GeV
are equal within errors.
At $\sqrt{s_{nn}}=17$GeV
   around  $m_T=0.5 $ the kaon phase densities
lie below those of the pions. However at  $\sqrt{s_{nn}}=5$GeV  this effect
is reversed.
Using Eqn.~\ref{eq:fkpi} we deduce from the drop of
$ \langle f_{k^+} \rangle /\langle f_{pi^+} \rangle$
  with $\sqrt{s_{nn}}$ that $\mu_u$ drops more rapidly with
 $\sqrt{s_{nn}}$ than   $\mu_s$.
Extrapolating the pion slope to   $m_T=1.05 $
we estimate that
$ \langle f_\pi  \rangle  >  \langle f_k \rangle \approx  \langle f_p \rangle. $
at   $\sqrt{s_{nn}}=17$GeV whereas  an extrapolation of the data at
$\sqrt{s_{nn}}=5$GeV would predict the opposite behavior.

\begin{figure}
\epsfig{file=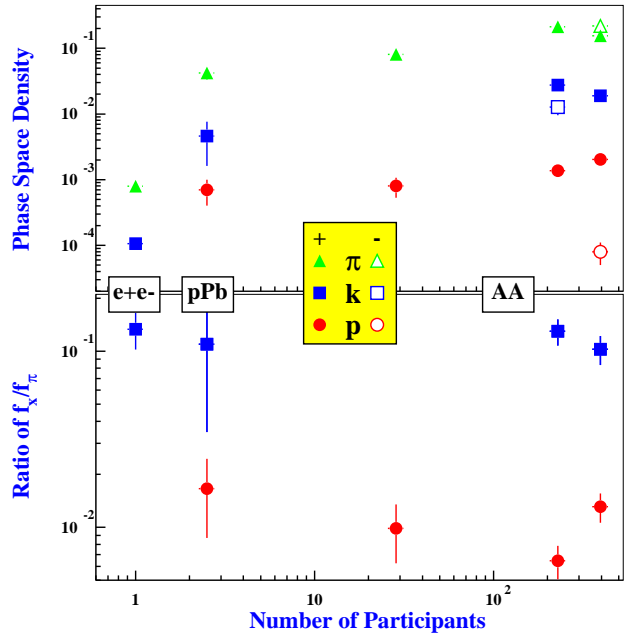,width=8.5cm,angle=0}
\caption{\label{fg:kfvnpart} Kaon, pion and proton phase space densities  (a)
 and ratios (b) versus the number of participants.}
\end{figure}

Figure~\ref{fg:kfvnpart} shows $\pi^\pm, k^\pm, p$ and $\bar p$
 phase space densities from
NA44 versus the number of participants. Also shown are $\pi^+$ and $k^+$
densities from $e^+e^-$ collisions at
 $\sqrt{s_{nn}}=91$GeV \cite{delphikk}.
   Although all phase densities increase markedly
with the number of participants their ratios are remarkably constant.
 A similar rise is in  $ \langle f_{k^+} \rangle$ is seen at the AGS where
  $ \langle f_{k^+} \rangle$  rises from Si+Au to Au+Au. %$0.016 \pm 0.003$ to
 For S+Pb
collisions NA44 has measured both $k^+$ and $k^-$ spectra and source sizes.
We find that      $ \langle f_{k^-} \rangle <  \langle f_{k^+} \rangle$. Therefore
from
Eqn.~\ref{eq:fkmdfkp} we deduce that $\mu_s < \mu_u$.

\section{Conclusions}

The kaon phase space $\langle f_{k^+} \rangle$  density drops with $m_T$
 and at a  $m_T$ is somewhat lower than  $\langle f_{\pi^+} \rangle$
 for $m_t \approx 0.5$ .
 This
 implies that $k^+$s have a larger chemical potential than pions.  Near
 $p_t=0$ we find
% $\langle f_{p} \rangle <\langle f_{k^+} \rangle <\langle f_{\pi^+}\rangle<1$.
 \[ \langle f_{\bar p} \rangle \ll \langle f_p \rangle \ll
 \langle f_{k^-} \rangle < \langle f_{k^+} \rangle  \ll
\langle f_{\pi^+} \rangle < \langle f_{\pi^-} \rangle < 1\]
% However because they have different slopes in $m_T$ the phase densities
% of pions, kaons and protons
% seem to become equal near $m_T=1.0$.
 However because  the protons have flatter  slopes in $m_T$ than the pions and kaons
 we would expect the proton phase space desnity to exceed that of
 the pions and kaons for t $m_T>1.2 GeV/c^2$.
We find that at low $p_T$ $\langle f_{k^+} \rangle$
and   $\langle f_{p} \rangle$
drops with   $\sqrt{s_{nn}}$. This presumably is because increase flow at
the higher energy overcomes the effect of extra particle production.
It will be interesting to see if this trend of dropping phase space density
continues up to  RHIC energies.
 As the number of participants increases
 $ \langle f_{k^+} \rangle$, $ \langle f_{\pi} \rangle$  and
 $ \langle f_p \rangle$  all increase but their ratios remain constant at SPS energies.

\section{Acknowledgments}
 I would like to thank the organizers for helping me attend the
 Strange Quarks in Matter Conference. This work was supported
 by the US DOE  Office of Science, grant  DE-FG03-93-ER40773.


\section{References}

\begin{thebibliography}{99}
\bibitem{BERTSCH} G. F. Bertsch, Phys. Rev. Lett. {\bf 72} 2349 (1994).; \\
{\it ibid.} {\bf 77} (1996) 789(E).
\bibitem{FERENC} D. Ferenc, U. Heinz, B. Tom\'a\v{s}ik, U.A. Wiedemann, and J.G. Cramer,
  Phys. Lett. B {\bf 457}, 347 (1999).
\bibitem{Murray} M. Murray and B. Holzer Phys. Rev. C. {\bf 63} 054901 (2001).
\bibitem{NA44KHbt}  I. G. Bearden {\it et al.},
   Phys. Rev Lett. {\bf 87} , 112301 (2001),
H. Beker    {\it et al.},  Z. Phys. C. {\bf 64}  209 (1994).
\bibitem{NA44ksng} I. G. Bearden {\it et al.}, Phys. Lett. B {\bf 471}, 6 (1999),
H. B\o ggild {\it et al.},   Phys. Rev. C  {\bf 59} 328 (1999).
\bibitem{WEID99A} U.A. Weidemann and U. Heinz,
 Phys. Rep {\bf 319} 145 (1999)  %nucl-th/9901094. %CERN-TH/99-15
\bibitem{CHAPMAN} S. Chapman, P. Scotto, and U. Heinz, Phys. Rev. Lett.
\bibitem{delphikk} For a review of LEP interferrometry results see,
O. Smirnova, ``Proceedings of the 20th International Symposium on Multiparticle
Dynamics, Tihany, October 2000,
edited by
T. Cs{\"o}rg\H o and W. Kittle  and published by World Scientific.
%\bibitem{NA44pbpi} I. G. Bearden {\it et al.},  Phys. Rev. C {\bf 58}, 1656 (1998).
\end{thebibliography}
\end{document}